\documentstyle[12pt]{article}
\catcode`@=11
\newcount\@tempcntc
\def\@citex[#1]#2{\if@filesw\immediate\write\@auxout{\string\citation{#2}}\fi
  \@tempcnta\z@\@tempcntb\m@ne\def\@citea{}\@cite{\@for\@citeb:=#2\do
    {\@ifundefined
       {b@\@citeb}{\@citeo\@tempcntb\m@ne\@citea\def\@citea{,}{\bf ?}\@warning
       {Citation `\@citeb' on page \thepage \space undefined}}%
    {\setbox\z@\hbox{\global\@tempcntc0\csname b@\@citeb\endcsname\relax}%
     \ifnum\@tempcntc=\z@ \@citeo\@tempcntb\m@ne
       \@citea\def\@citea{,}\hbox{\csname b@\@citeb\endcsname}%
     \else
      \advance\@tempcntb\@ne
      \ifnum\@tempcntb=\@tempcntc
      \else\advance\@tempcntb\m@ne\@citeo
      \@tempcnta\@tempcntc\@tempcntb\@tempcntc\fi\fi}}\@citeo}{#1}}
\def\@citeo{\ifnum\@tempcnta>\@tempcntb\else\@citea\def\@citea{,}%
  \ifnum\@tempcnta=\@tempcntb\the\@tempcnta\else
   {\advance\@tempcnta\@ne\ifnum\@tempcnta=\@tempcntb \else \def\@citea{--}\fi
    \advance\@tempcnta\m@ne\the\@tempcnta\@citea\the\@tempcntb}\fi\fi}
\catcode`@=12
\begin{document}
\title{\vskip-3cm{\baselineskip14pt
\centerline{\normalsize MPI/PhT/97--010\hfill}
\centerline{\normalsize NYU--TH--97/03/01\hfill}
\centerline{\normalsize hep--ph/9703226\hfill}
\centerline{\normalsize February 1997\hfill}}
\vskip1.5cm
Estimations of Order $\alpha_s^3$ and $\alpha_s^4$ Corrections to
Mass-Dependent Observables}
\author{K.G. Chetyrkin$^{1,2}$, B.A. Kniehl$^2$, A. Sirlin$^3$\\
$^1$ Institute for Nuclear Research, Russian Academy of Sciences,\\
60th October Anniversary Prospect 7a, Moscow 117312, Russia\\
$^2$ Max-Planck-Institut f\"ur Physik (Werner-Heisenberg-Institut),\\
F\"ohringer Ring 6, 80805 Munich, Germany\\
$^3$ Department of Physics, New York University,\\
4 Washington Place, New York, NY 10003, USA}
\date{}
\maketitle
\begin{abstract}
A simple procedure to estimate ${\cal O}(\alpha_s^3)$ and
${\cal O}(\alpha_s^4)$ corrections to mass-dependent observables is
conjectured.
The method is tested in a number of cases where the ${\cal O}(\alpha_s^3)$
contribution is exactly known, and reasonable agreement is found.

\medskip
\noindent
PACS numbers: 11.15.Me, 11.25.Db, 12.38.-t, 12.38.Cy
\end{abstract}
\newpage

Given the great difficulty of evaluating higher-order corrections, it is very
desirable to have reasonable methods to estimate their sign and magnitude.
In fact, significant and interesting investigations in this subject have been
carried out in the past \cite{gru,ste,kub,kat,bro,sur}.
The aim of this note is to propose a simple estimation method to treat an
important class of mass-dependent observables.

We first recall salient features of the estimation methods proposed in the
literature.
Calling $R(s)$ an observable depending on a single time-like kinematic
variable $s$, such as a squared centre-of-mass energy, we consider the QCD
expansion
\begin{equation}
R(s)=\sum_{n=0}^\infty R_n(s,\mu^2)a^n(\mu^2),
\label{A1}
\end{equation}
where $a(\mu^2)=\alpha_s^{(n_f)}(\mu^2)/\pi$, $\mu$ is the renormalization 
scale, and $n_f$ is the number of flavours active at that scale.
By factoring out an appropriate power of $s$, it is always possible to render
$R(s)$ dimensionless.
Henceforth, we shall adopt this convention.
As $R(s)$ is renormalization-group invariant, we may choose $\mu^2=s$, in
which case Eq.~(\ref{A1}) becomes
\begin{equation}
R(s)=\sum_{n=0}^\infty r_na^n(s), 
\label{A2}
\end{equation}
where $r_n=R_n(s,s)$.
If $R(s)$ does not depend on masses or other kinematical variables, the $r_n$
are numerical constants.
Estimations of higher-order corrections using optimization procedures based on
the fastest apparent convergence (FAC) \cite{gru} and the principle of minimal
sensitivity (PMS) \cite{ste} have been carried out in two main scenarios:

\noindent
(i) Suppose that $r_0$, $r_1$, and $r_2$ are known. 
Then the FAC and PMS approaches lead to the estimates
\cite{gru,ste}
\begin{eqnarray}
r_3^{\rm FAC}&=&r_2\left(\frac{r_2}{r_1}+\frac{\beta_1}{\beta_0}\right),
\nonumber\\
r_3^{\rm PMS}&=&r_1\left(\frac{r_2}{r_1}+\frac{\beta_1}{2\beta_0}\right)^2,
\label{A3}
\end{eqnarray}
where
\begin{equation}
\beta_0=\frac{1}{4}\left(11-\frac{2}{3}n_f\right),
\quad
\beta_1=\frac{1}{16}\left(102-\frac{38}{3}n_f\right)
\end{equation}
are the first two coefficients of the QCD $\beta$ function. 

\noindent
(ii) 
Suppose that $r_0$, $r_1$, $r_2$, and $r_3$ are known.
Then in both the FAC and PMS methods one finds \cite{gru,ste,kub,kat}
\begin{equation}
r_4^{\rm FAC/PMS}=r_2\left(3\frac{r_3}{r_1}-2\frac{r_2^2}{r_1^2} 
-\frac{r_2\beta_1}{2r_1\beta_0}+\frac{\beta_2}{\beta_0}\right),
\label{A6}
\end{equation}
where, in the $\overline{\rm MS}$ scheme, \cite{tar}
\begin{equation}
\beta_2=\frac{1}{64}\left(\frac{2857}{2}-\frac{5033}{18}n_f
+\frac{325}{54}n_f^2\right).
\end{equation}  
If $r_3$ is not known, we may consider employing the estimates from
Eq.~(\ref{A3}).
Inserting these expressions into Eq.~(\ref{A6}), one finds
\begin{eqnarray}
r_4^{\rm FAC}&=&r_2\left(\frac{r_2^2}{r_1^2}+\frac{5r_2\beta_1}{2r_1\beta_0} 
+\frac{\beta_2}{\beta_0}\right),
\nonumber\\
r_4^{\rm PMS}&=&r_2\left(\frac{r_2^2}{r_1^2}+\frac{5r_2\beta_1}{2r_1\beta_0} 
+\frac{\beta_2}{\beta_0}+\frac{3\beta_1^2}{4\beta_0^2}\right).
\label{A8}
\end{eqnarray}

Formally, the FAC and PMS procedures could be applied in both Minkowskian and
Euclidean spaces.
However, on general grounds it is expected that the optimization procedures
are more accurate when applied in Euclidean space, as one avoids the presence
of physical thresholds.
Accordingly, when the theoretical expansion for an observable is given 
Minkowskian space, it has been proposed \cite{kat} to apply the optimization
procedures to an associated function defined in the Euclidean region, namely
\begin{equation}
D(Q^2)=Q^2\int_0^\infty ds\frac{R(s)}{(s+Q^2)^2},
\label{A9}
\end{equation}
with $Q^2\ge0$.
$R(s)$ admits the inverse integral representation
\begin{equation}
R(s)=\frac{1}{2\pi i}\int_{-s-i\varepsilon}^{-s+i\varepsilon}ds^\prime
\frac{D(s^\prime)}{s^\prime},
\end{equation}
where $D(s^\prime)$ is the analytic continuation of $D(Q^2)$ to the complex
plane.
Inserting the expansion of Eq.~(\ref{A1}) into Eq.~(\ref{A9}), carrying out
the integration, and then setting $\mu^2=Q^2$, one finds
\begin{equation}
D(Q^2)=\sum_{n=0}^\infty d_na^n(Q^2),
\label{A12}
\end{equation}
where
\begin{eqnarray}
d_0&=&r_0,\quad
d_1=r_1,\quad
d_2=r_2,
\label{AA13}\\
d_3&=&r_3+\frac{\pi^2}{3}r_1\beta_0^2,\quad
d_4=r_4+\pi^2\beta_0\left(r_2\beta_0+\frac{5}{6}r_1\beta_1\right).
\label{A13}
\end{eqnarray}
In the Euclidean approach, one estimates $d_3$ and $d_4$ using the expressions
analogous to Eqs.~(\ref{A3}), (\ref{A6}), and (\ref{A8}) with $r_n$ replaced 
by $d_n$, and obtains $r_3$ and $r_4$ from Eq.~(\ref{A13}).

At this point, we turn our attention to mass-dependent observables of the form 
\begin{equation}
T(s)=m^2(\mu^2)\sum_{n=0}^\infty T_n(s,\mu^2)a^n(\mu^2),
\label{A14}
\end{equation}
where $m(\mu^2)$ is a running quark mass.
Again, without loss of generality, we may assume that the $T_n(s,\mu^2)$ are
dimensionless.
Setting $\mu^2=s$ and defining $t_n=T_n(s,s)$, we have
\begin{equation}
T(s)=m^2(s)\sum_{n=0}^\infty t_na^n(s).
\label{A15}
\end{equation}
If $T(s)/m^2(s)$ does not depend on masses or other kinematic variables, the
$t_n$ are numerical constants. 
The associated function defined in the Euclidean region is
\begin{equation}
F(Q^2)=Q^2\int_0^\infty ds\frac{T(s)}{(s+Q^2)^2}.
\label{A16}
\end{equation}
In order to carry out the $s$ integration in Eq.~(\ref{A16}), we substitute in
Eq.~(\ref{A15}) the expansions~\cite{kni}
\begin{eqnarray}
a(s)&=&a(\mu^2)\left\{1-a(\mu^2)\beta_0\ell
+a^2(\mu^2)\ell(\beta_0^2\ell-\beta_1)
+a^3(\mu^2)\ell\left(-\beta_0^3\ell^2+\frac{5}{2}\beta_0\beta_1\ell-\beta_2
\right)\right.
\nonumber\\
&+&\left.
a^4(\mu^2)\ell\left[\beta_0^4\ell^3-\frac{13}{3}\beta_0^2\beta_1\ell^2
+3\left(\frac{\beta_1^2}{2}+\beta_0\beta_2\right)\ell-\beta_3\right]
+{\cal O}(a^5\ell^5)\right\},
\label{A17}\\
m(s)&=&m(\mu^2)\left\{1-a(\mu^2)\gamma_0\ell
+a^2(\mu^2)\ell\left[\frac{\gamma_0}{2}(\beta_0+\gamma_0)\ell-\gamma_1\right]
\right.
\nonumber\\
&+&a^3(\mu^2)\ell\left[-\frac{\gamma_0}{3}(\beta_0+\gamma_0)
\left(\beta_0+\frac{\gamma_0}{2}\right)\ell^2+\left(\frac{\beta_1\gamma_0}{2}
+\gamma_1(\beta_0+\gamma_0)\right)\ell-\gamma_2\right]
\nonumber\\
&+&a^4(\mu^2)\ell\left[\frac{\gamma_0}{4}(\beta_0+\gamma_0)
\left(\beta_0+\frac{\gamma_0}{2}\right)\left(\beta_0+\frac{\gamma_0}{3}\right)
\ell^3\right.\nonumber\\
&-&\left(\frac{\beta_1\gamma_0}{2}\left(\frac{5}{3}\beta_0+\gamma_0\right)
+\gamma_1(\beta_0+\gamma_0)\left(\beta_0+\frac{\gamma_0}{2}\right)\right)
\ell^2
\nonumber\\
&+&\left.\left.\left(\frac{\beta_2\gamma_0}{2}+\gamma_1\left(\beta_1
+\frac{\gamma_1}{2}\right)+\gamma_2\left(\frac{3}{2}\beta_0+\gamma_0\right)
\right)\ell-\gamma_3\right]+{\cal O}(a^5\ell^5)\right\},
\end{eqnarray}
where $\ell=\ln(s/\mu^2)$ and, in the $\overline{\rm MS}$ scheme, the
coefficients of the mass anomalous dimension are \cite{lar}
\begin{eqnarray}
\gamma_0&=&1,\quad
\gamma_1=\frac{1}{16}\left(\frac{202}{3}-\frac{20}{9}n_f\right),
\nonumber\\
\gamma_2&=&\frac{1}{64}\left\{1249
-\left[\frac{2216}{27}+\frac{160}{3}\zeta(3)\right]n_f
-\frac{140}{81}n_f^2\right\}.
\end{eqnarray}
Here, $\zeta$ is Riemann's zeta function, with value $\zeta(3)\approx1.20206$.
The coefficient $\gamma_3$ is presently unknown.
For completeness, we have also presented the ${\cal O}(a^4)$ term in 
Eq.~(\ref{A17}), although we shall not need it here.
These substitutions bring Eq.~(\ref{A15}) into the form of Eq.~(\ref{A14}) with
coefficient functions $T_n(s,\mu^2)$ that depend on $s$ only through powers of
$\ell$. 
Inserting the expression thus obtained into Eq.~(\ref{A16}) and using the
elementary integrals
\begin{equation}
Q^2\int_0^\infty ds\frac{\{1;\ell;\ell^2;\ell^3;\ell^4\}}{(s+Q^2)^2} 
=\left\{1;L;L^2+\frac{\pi^2}{3};L^3+\pi^2L;
L^4+2\pi^2L^2+\frac{7\pi^4}{15}\right\},
\end{equation}
where $L=\ln(Q^2/\mu^2)$, one obtains an expansion of the form 
\begin{equation}
F(Q^2)=m^2(\mu^2)\sum_{n=0}^{\infty}F_n(Q^2,\mu^2)a^n(\mu^2).
\label{A21}
\end{equation}
Setting $\mu^2=Q^2$, Eq.~(\ref{A21}) becomes 
\begin{equation}
F(Q^2)=m^2(Q^2)\sum_{n=0}^\infty f_na^n(Q^2),
\label{A22}
\end{equation}
where $f_n=F_n(Q^2,Q^2)$ are numerical constants.
Specifically, one finds
\begin{eqnarray}
f_0&=&t_0,\quad
f_1=t_1,\quad
f_2=t_2+\frac{\pi^2}{3}t_0\gamma_0(\beta_0+2\gamma_0),
\label{A23}\\
f_3&=&t_3+\frac{\pi^2}{3}\{t_1(\beta_0+\gamma_0)(\beta_0+2\gamma_0)
+t_0[\beta_1\gamma_0+2\gamma_1(\beta_0+2\gamma_0)]\},
\label{A24}\\
f_4&=&t_4+\pi^2\left\{t_2(\beta_0+\gamma_0)\left(\beta_0+\frac{2}{3}\gamma_0
\right)+t_1\left[\beta_1\left(\frac{5}{6}\beta_0+\gamma_0\right)
+\frac{4}{3}\gamma_1(\beta_0+\gamma_0)\right]\right.
\nonumber\\
&+&\left.t_0\left[\frac{\beta_2\gamma_0}{3}+\frac{2}{3}\gamma_1(\beta_1
+\gamma_1)+\gamma_2\left(\beta_0+\frac{4}{3}\gamma_0\right)\right]\right\}
\nonumber\\
&+&\frac{7\pi^4}{15}t_0\gamma_0(\beta_0+\gamma_0)(\beta_0+2\gamma_0)
\left(\frac{\beta_0}{2}+\frac{\gamma_0}{3}\right).
\label{A25}
\end{eqnarray}
The case of Eqs.~(\ref{A1}) and (\ref{A2}), in which 
the $m^2(\mu^2)$ factor is not present, can be obtained
from Eqs.~(\ref{A23})--(\ref{A25}) by setting $\gamma_i=0$
($i=0,1,2$).
In fact, the relations between the $f_n$ and $t_n$ then become identical to
those between the $d_n$ and $r_n$ in Eqs.~(\ref{AA13}) and (\ref{A13}).

Our proposal is to apply the estimation procedure described before to the
$f_n$ expansion of Eq.~(\ref{A22}) and to obtain the corresponding $t_n$
coefficients via Eqs.~(\ref{A24}) and (\ref{A25}).
For instance, if the $t_n$ are known for $n\le2$, we obtain $f_n$ for $n\le2$
from Eq.~(\ref{A23}) and estimate $f_3$ using the expressions analogous to
Eq.~(\ref{A3}) with $r_n$ replaced by $f_n$.
The estimate for $t_3$ is then obtained from Eq.~(\ref{A24}).
If the $t_n$ are known for $n\le 3$, we obtain $f_n$ for $n\le 3$ from
Eqs.~(\ref{A23}) and (\ref{A24}), $f_4$ is estimated from the expression
analogous to Eq~(\ref{A6}) with $r_n$ replaced by $f_n$, and $t_4$ follows
from Eq.~(\ref{A25}).
If $t_3$ is not known, we may also attempt to estimate $f_4$ from the
expressions analogous to Eq.~(\ref{A8}), and $f_4$ once more from
Eq.~(\ref{A25}).
This proposal essentially relates the estimation of the higher-order
coefficients in the mass-dependent expansion to the previously considered
mass-independent case.
It should be pointed out that there is an element of arbitrariness in this
approach.
In Eq.~(\ref{A21}), we have set $\mu^2=Q^2$ and proposed to apply the
optimization procedure to $\sum_{n=0}^\infty f_na^n(Q^2)$, the cofactor of
$m^2(Q^2)$.
Had we chosen a different scale $\mu^2\ne Q^2$ in Eq.~(\ref{A21}), the
expansion in Eq.~(\ref{A22}) and the estimation procedure would be different.
On the other hand, the choice $\mu^2=Q^2$ seems natural and convenient, as all
the logarithms in the $F_n(Q^2,\mu^2)$ vanish.
In fact, this feature has an additional very useful property: it renders the
analysis of $t_4$ independent of the unknown coefficient $\gamma_3$. 

\bigskip

\begin{table}[ht]
\renewcommand{\arraystretch}{1.3}
\begin{center}
\begin{tabular}{|c|c|c|c|c|}
\hline
$n_f$ & $t_3^{\rm exact}$ & $t_3^{\rm FAC}$ & $t_3^{\rm PMS}$ &
$t_4^{\rm FAC/PMS}$ \\ 
\hline\hline
3 & 89.156 & 75.729 & 80.206 & $-945.28$\\ 
\hline
4 & 65.198 & 64.956 & 68.316 & $-1098.8$\\ 
\hline
5 & 41.758 & 53.295 & 55.547 & $-1237.4$\\ 
\hline
\end{tabular}
\end{center}
\caption{\label{t1}
Estimations of $t_3$ and $t_4$ in $\Gamma(H\to q\bar q)$ based on the FAC and
PMS optimizations of the associated function $F(Q^2)/m^2(Q^2)$, defined in the
Euclidean region, and Eqs.~(\ref{A23})--(\ref{A25}).
The estimation of $t_4$ employs the exact value of $t_3$.}
\end{table}

\bigskip

\begin{table}[ht]
\renewcommand{\arraystretch}{1.3}
\begin{center}
\begin{tabular}{|c|c|c|c|c|}
\hline
$n_f$ & $t_3^{\rm exact}$ & $t_3^{\rm FAC}$ & $t_3^{\rm PMS}$ &
$t_4^{\rm FAC/PMS}$ \\ 
\hline
\hline
3 & 89.156 & 235.82 & 240.30 & $-527.81$\\
\hline
4 & 65.198 & 211.20 & 214.56 & $-748.62$\\
\hline
5 & 41.758 & 186.67 & 188.92 & $-949.39$\\
\hline
\end{tabular}
\end{center}
\caption{\label{t2}
As in Table~\ref{t1}, but using the original function $T(s)/m^2(s)$, defined
in the Minkowskian region.}
\end{table}

\bigskip

An interesting application is the estimation of the ${\cal O}(\alpha_s^3)$
and ${\cal O}(\alpha_s^4)$ coefficients in the evaluation of the partial width
$\Gamma(H\to{\rm hadrons})$ involving final-state quarks with running mass
$m(\mu^2)\ll M_H$.
The relevant expansion \cite{che} is of the form of Eq.~(\ref{A15}) with
$\sqrt s=M_H$, $t_0=1$, $t_1=17/3$, and $t_2\approx35.93996-1.35865\,n_f$,
where $n_f$ is the number of active flavours at scale $\sqrt s=M_H$.
The calculation assumes that there is one massive flavour, identical with that
present in the final state of the reaction $H\to q\bar q$, and $n_f-1$
massless ones. 
Table~\ref{t1} compares the $t_3$ estimates of the proposed procedure, based
on the FAC and PMS optimizations of $F(Q^2)/m^2(Q^2)$, with the exact result
and also provides the $t_4$ estimate obtained using the exact value of $t_3$.
Table~\ref{t2} displays the corresponding estimations based on the
optimization of the original function $T(s)/m^2(s)$ defined in the Minkowskian
domain.
We see that the predicted signs for $t_3$ are correct, but it is apparent that
the magnitude of the estimations is much closer to the exact result in the
Euclidean approach.
In fact, the estimations of $t_3$ in Table~\ref{t1} are fairly good:
$t_3^{\rm FAC}$ shows errors of $(-15,-0.4,+28)\%$ for $n_f=3,4,5$,
respectively;
for $t_3^{\rm PMS}$ the corresponding errors are $(-10,+5,+33)\%$. 
We also see that the $t_4$ coefficients are predicted to be large and
negative. 

\bigskip

\begin{table}[ht]
\renewcommand{\arraystretch}{1.3}
\begin{center}
\begin{tabular}{|c|c|c|c|c|}
\hline
$n_f$ & $t_3^{\rm exact}$ & $t_3^{\rm FAC}$ & $t_3^{\rm PMS}$ &
$t_4^{\rm FAC/PMS}$ \\ 
\hline\hline
3 & $-87.394$ & $-105.06$ & $-103.75$ & $-582.03$\\
\hline
4 & $-83.356$ & $-98.597$ & $-97.609$ & $-487.93$\\
\hline
5 & $-79.598$ & $-92.475$ & $-91.813$ & $-400.78$\\
\hline
\end{tabular}
\end{center}
\caption{\label{t3}
Estimations of $t_3$ and $t_4$ in the $m_c^2$ contribution to
$\Gamma(W^+\to c\bar s)$ based on the FAC and PMS optimizations of the
associated function $F(Q^2)/m^2(Q^2)$, defined in the Euclidean region, and
Eqs.~(\ref{A23})--(\ref{A25}).
The estimation of $t_4$ employs the exact value of $t_3$.}
\end{table}

\bigskip

\begin{table}[ht]
\renewcommand{\arraystretch}{1.3}
\begin{center}
\begin{tabular}{|c|c|c|c|c|}
\hline
$n_f$ & $t_3^{\rm exact}$ & $t_3^{\rm FAC}$ & $t_3^{\rm PMS}$ &
$t_4^{\rm FAC/PMS}$ \\ 
\hline\hline
3 & 32.096 & 8.6895 & 11.587 & $-331.80$\\
\hline
4 & 20.311 & 4.2754 & 6.4494 & $-400.40$\\
\hline
5 & 8.6525 & $-0.70475$ & 0.75256 & $-464.41$\\
\hline
\end{tabular}
\end{center}
\caption{\label{t4}
As in Table~\protect\ref{t3}, but for $Z\to b\bar b$.}
\end{table}

\bigskip

There are two other cases in which $t_3$ is exactly known: 
these are the terms proportional to $m_q^2$ in the absorptive parts of the
axial-vector correlators pertinent to the parton-level decays 
$W^+\to c\bar s$ and $Z\to b\bar b$ \cite{kue}.
Here, $m_c$ is evaluated with $n_f=4$ at $\sqrt s=M_W$ and $m_b$ with $n_f=5$
at $\sqrt s=M_Z$.
In either case, it is assumed that the remaining $n_f-1$ quarks are massless.
The corresponding coefficients are $t_1=5/3$,
$t_2\approx-3.06004-0.02532\,n_f$ for $W^+\to c\bar s$ and $t_1=11/3$,
$t_2\approx18.02329-0.74754\,n_f$ for $Z\to b\bar b$.
Tables~\ref{t3} and \ref{t4} show the FAC and PMS estimations for these
transitions in the Euclidean approach.
For illustration, we also consider other values of $n_f$.
In the case of $W^+\to c\bar s$, we see once more that the $t_3$ estimations
are fairly good, with relative errors of 20\% or below.
Instead, in the case of $Z\to b\bar b$, the relative errors are large.
We note, however, that in this case both the exact and estimated $t_3$ values
are relatively small.
In fact, a simple, way to characterize Tables~\ref{t1}, \ref{t3}, and \ref{t4}
is to say that the $t_3$ estimations have absolute errors of order 20 or
below.
When $t_3^{\rm exact}$ is large, as in Tables~\ref{t1} and \ref{t3}, this
leads to fairly accurate results. 

\bigskip

\begin{table}[ht]
\renewcommand{\arraystretch}{1.3}
\begin{center}
\begin{tabular}{|c|c|c|c|c|}
\hline
$n_f$ & $t_3^{\rm FAC}$ & $t_3^{\rm PMS}$ & $t_4^{\rm FAC}$ & $t_4^{\rm PMS}$
\\ 
\hline\hline
3 & 152.71 & 153.76 & 2083.8 & 2123.4\\
\hline
4 & 124.10 & 124.89 & 1544.1 & 1571.4\\
\hline
5 & 97.729 & 98.259 & 1091.0 & 1107.8\\
\hline
6 & 73.616 & 73.903 & 718.74 & 727.00\\
\hline
\end{tabular}
\end{center}
\caption{\label{t5}
Estimations of $t_3$ and $t_4$ in Eq.~(\ref{A26}) based on the FAC and PMS
optimizations of the associated function $F(Q^2)/m(Q^2)$, defined in the
Euclidean region, and Eqs.~(\ref{A28})--(\ref{A30}).}
\end{table}

\bigskip

There are some important cases in which the $t_3$ coefficients are not known.
Examples include mass relations of the form
\begin{equation}
M_q=\mu_q\sum_{n=0}^{\infty}t_na^n(\mu_q^2),
\label{A26}
\end{equation}
where $M_q$ is the pole mass and $\mu_q=m_q(\mu_q^2)$ is the
$\overline{\rm MS}$ mass of quark $q$.
In contrast to the previous applications, Eq.~(\ref{A26}) does not depend on
an external mass parameter such as $M_H$, $M_W$, or $M_Z$ which, in principle,
can have arbitrary values independent of $m_q$.
On the other hand, defining $T(s)=m_q(s)\sum_{n=0}^{\infty}t_na^n(s)$ for
arbitrary $s\ge0$, we have the mathematical identity
\begin{equation}
M_q=\frac{1}{2\pi i}\int_{-\mu_q-i\varepsilon}^{-\mu_q+i\varepsilon}ds^\prime
\int_0^\infty ds\frac{T(s)}{(s + s^\prime)^2}.
\end{equation}
In analogy with the previous applications, we introduce the function $F(Q^2)$
defined by Eq.~(\ref{A16}) in the Euclidean region $Q^2\ge0$. 
Because of their linear dependence on $m_q$, the relations 
between the $f_n$ and $t_n$ are different from those in
Eqs.~(\ref{A23})--(\ref{A25}).
We now have
\begin{eqnarray}
f_0&=&t_0,\quad f_1=t_1,\quad
f_2=t_2+\frac{\pi^2}{6}t_0\gamma_0(\beta_0+\gamma_0),
\label{A28}\\
f_3&=&t_3+\frac{\pi^2}{3}\left\{t_1(\beta_0+\gamma_0)
\left(\beta_0+\frac{\gamma_0}{2}\right)
+t_0\left[\frac{\beta_1\gamma_0}{2}+\gamma_1(\beta_0+\gamma_0)\right]\right\},
\label{A29}\\
f_4&=&t_4
+\pi^2\left\{t_2\left(\beta_0+\frac{\gamma_0}{2}\right)
\left(\beta_0+\frac{\gamma_0}{3}\right)
+t_1\left[\frac{\beta_1}{2}\left(\frac{5}{3}\beta_0+\gamma_0\right)
+\frac{\gamma_1}{3}(2\beta_0+\gamma_0)\right]\right.
\nonumber\\
&+&\left.t_0\left[\frac{\beta_2\gamma_0}{6}
+\frac{\gamma_1}{3}\left(\beta_1+\frac{\gamma_1}{2}\right)
+\gamma_2\left(\frac{\beta_0}{2}+\frac{\gamma_0}{3}\right)\right]\right\}
\nonumber\\
&+&\frac{7\pi^4}{60}t_0\gamma_0(\beta_0+\gamma_0)
\left(\beta_0+\frac{\gamma_0}{2}\right)
\left(\beta_0+\frac{\gamma_0}{3}\right).
\label{A30}
\end{eqnarray}
Assuming once more that $n_f-1$ quarks are massless, we have in the case of
Eq.~(\ref{A26}) $t_0=1$, $t_1=4/3$, and $t_2\approx14.48476-1.04137\,n_f$
\cite{gra}.
Table~\ref{t5} displays the coefficients $t_3$ and $t_4$ in Eq.~(\ref{A26})
estimated by the FAC and PMS optimizations of the associated function
$F(Q^2)/m(Q^2)$, using Eqs.~(\ref{A3}) and (\ref{A8}) with $r_n$
replaced by $f_n$.
The entry for $n_f=6$ corresponding to $M_t/\mu_t$, where $t$ is the top
quark, gives an expansion very close to that obtained in a different approach 
based on an optimization of the ratios $M_t/m_t(M_t^2)$ and $m_t(M_t^2)/\mu_t$
\cite{sir}.
For example, if the expansion is made in powers of $a(M_t^2)$ rather than
$a(\mu_t^2)$, the FAC result in Table~\ref{t5} becomes
\begin{eqnarray}
M_t&=&\mu_t\left[1+\frac{4}{3}a(M_t^2)+8.2366\,a^2(M_t^2)+79.838\,a^3(M_t^2)
+835.69\,a^4(M_t^2)\right.
\nonumber\\
&+&\left.{\cal O}(a^5)\right],
\label{A31}
\end{eqnarray}
while the approach of Ref.~\cite{sir} leads to 
\begin{eqnarray}
M_t&=&\mu_t\left[1+\frac{4}{3}a(M_t^2)+8.2366\,a^2(M_t^2)+76.172\,a^3(M_t^2)
+797.95\,a^4(M_t^2)\right.
\nonumber\\
&+&\left.{\cal O}(a^5)\right].
\label{A32}
\end{eqnarray}

It is well known that, if expressed in terms of $\mu_t$, the QCD corrections
to $\Delta\rho_f$, the fermionic contributions to the electroweak $\rho$
parameter, are of the form \cite{avd}
\begin{equation}
\Delta\rho_f=\frac{3G_\mu\mu_t^2}{8\pi^2\sqrt2}
\left[1-0.19325\,a(M_t^2)-3.9696\,a^2(M_t^2)+{\cal O}(a^3)\right],
\label{A33}
\end{equation}
where $G_\mu$ is Fermi's constant.
Most of the second-order coefficient in Eq.~(\ref{A33}), $-4.2072$, arises
from the opening of a new channel, namely the double-triangle diagram.
It is clear that at present there is no sufficient information to optimize
Eq.~(\ref{A33}).
However, if one makes the reasonable assumption that the higher-order terms in
Eq.~(\ref{A33}) follow the pattern of rather small coefficients shown by the
leading contributions, we can combine this result with Eqs.~(\ref{A31}) or
(\ref{A32}) to estimate $t_3$ and $t_4$ in $\Delta\rho_f/M_t^2$.
Substituting Eq.~(\ref{A31}) into Eq.~(\ref{A33}), one finds
\begin{eqnarray}
\Delta\rho_f&=&\frac{3G_\mu M_t^2}{8\pi^2\sqrt2}
\left[1-2.8599\,a(M_t^2)-14.594\,a^2(M_t^2)-90.527\,a^3(M_t^2)\right.
\nonumber\\
&-&\left.924.88\,a^4(M_t^2)+{\cal O}(a^5)\right].
\label{A34}
\end{eqnarray}
The pattern of rapidly increasing coefficients of the same sign displayed in
Eqs.~(\ref{A31}), (\ref{A32}), and (\ref{A34}) also emerges from the analysis
of infrared-renormalon contributions \cite{ben}.
As pointed out in Ref.~\cite{sir}, expansions with much better convergence
properties can be obtained by optimizing the scale $\mu$ at which $a(\mu^2)$
is evaluated.

In summary, we have presented a procedure to estimate ${\cal O}(\alpha_s^3)$
and ${\cal O}(\alpha_s^4)$ corrections to a class of mass-dependent
observables of the type shown in Eqs.~(\ref{A14}) and (\ref{A15}).
Although there are elements of arbitrariness in its construction, the proposed
algorithm, based on the optimization of associated expansions in the Euclidean
region, is quite simple and obviates the dependence on the unknown coefficient
$\gamma_3$ of the mass anomalous dimension.
In the cases where the $t_3$ coefficients are exactly known, the proposed
algorithm estimates $t_3$ with absolute errors of order 20 or below.
In two of the three cases considered, $t_3^{\rm exact}$ is large, and the
estimations are fairly accurate, with reasonable relative errors.
We have then generalized the estimation algorithm to important expansions of
the form of Eq.~(\ref{A26}), where the $t_3$ coefficients are so far unknown.
The corresponding $t_3$ and $t_4$ estimations turn out to be close to those
found in recent analyses based on alternative optimizations procedures.

One of us (A.S.) would like to thank the Max Planck Institute in Munich and
the Benasque Center for Physics in Benasque, Spain, for their kind hospitality
during the summer of 1996, when part of this work was done.
His work was supported in part by NSF Grant PHY--9313781.

\end{document}